\documentstyle[12pt]{article}
\begin{document}
\begin{flushright}
hep-th/0106215\\
SNBNCBS-2001
\end{flushright}

\vskip 2cm

\begin{center}

{\bf \Large {Superfield approach to  BRST cohomology}}

\vskip 3.5cm

{\bf R. P. Malik} 
\footnote{ E-mail address: malik@boson.bose.res.in }\\
{\it S. N. Bose National Centre for Basic Sciences,}\\
{\it Block-JD, Sector-III, Salt Lake, Calcutta- 700 098, India}\\

\vskip 4.0cm

\end{center}

\noindent
{\bf Abstract:}
In the framework of superfield formalism, we discuss some aspects of the 
cohomological features of a two ($1+1$)-dimensional free Abelian gauge theory 
described by a Becchi-Rouet-Stora-Tyutin (BRST) invariant Lagrangian 
density.  We demonstrate that the conserved and nilpotent (anti-)BRST- and 
(anti-)co-BRST charges are the generators of translations along the 
Grassmannian directions of the four ($2+2$)-dimensional supermanifold. A 
bosonic symmetry is shown to be generated by a Noether conserved charge
that generates a translation along a bosonic direction of the supermanifold 
which turns out to be equivalent to a couple of successive translations along 
the two different and independent Grassmannian directions of the same 
supermanifold. Algebraically, these charges are found to be 
analogous to the de Rham cohomology operators of differential geometry.

\baselineskip=16pt

\vskip 1cm
%\noindent PACS Number(s) : 03.70; 05.30; 02.20; 71.27
%\thispagestyle{empty}
\newpage

\pagestyle{headings}

\noindent 
{\bf 1 Introduction}\\

\noindent
The existence of some new local, covariant and continuous symmetries
and their possible connections with the mathematics of differential geometry,
cohomology and Hodge decomposition theorem (HDT) for the 
two ($1+1$)-dimensional (2D) free- as well as interacting (non-)Abelian gauge 
theories  have been recently established in a set of papers [1--5]. The 
topological nature of the 2D free Abelian- and  self-interacting non-Abelian 
gauge theories (having no interaction with matter fields) has also been 
demonstrated
by exploiting the conserved and nilpotent (anti-)BRST charges $(Q_{(a)b})$, 
(anti-)dual BRST charges $(Q_{(a)d})$, a conserved ghost charge $Q_{g}$ and a 
bosonic charge $ Q_{w} = \{ Q_{d}, Q_{b} \} = \{ Q_{ab}, Q_{ad} \}$ which 
generate some interesting local, covariant and continuous  
symmetry transformations for the 2D BRST invariant Lagrangian
density [6]. These local conserved charges have also been utilized to express
HDT  in the quantum Hilbert space of states where any arbitrary state
$| \Psi>_{n} $ with the ghost number $n$,
(i.e. $i Q_{g} |\Psi>_{n} = n |\Psi>_{n}$) can be written as the sum of a
harmonic state $|\omega>_{n}$ ($ Q_{w} |\omega>_{n} = 0, Q_{b} |\omega>_{n} = 0,
Q_{d} |\omega>_{n} = 0$), a BRST exact state $ Q_{b} |\xi>_{n-1}$ and a 
co-BRST exact state $ Q_{d} |\chi>_{n + 1}$ as
\footnote{This equation is the analogue of the HDT which states that
any arbitrary $n$-form $f_{n}$ ($ n = 0, 1, 2,....$),
on a compact manifold, can be decomposed into a
harmonic form $\omega_{n}$ $( \Delta \omega_{n} = 0, d 
\omega_{n} = 0, \delta \omega_{n} = 0)$, an exact form 
$ d\; g_{n-1}$ and a co-exact form $ \delta\; h_{n+1}$ as:
$f_{n} = \omega_{n} + d\; g_{n-1} + \delta\; h_{n+1}$
where $\delta (= \pm\; {*}\; d \;{*})$ is the Hodge dual of $d$ 
(with $ d^2 = 0,\delta^2 = 0$ ; $* =$  Hodge duality operation)
and Laplacian $\Delta$ is defined as $ \Delta = ( d + \delta )^2
= d \delta + \delta d $ [7-10]. The set $( d, \delta, \Delta)$ is known
as the de Rham cohomology operators of differential geometry and
they obey the algebra: $d^2 = \delta^2 = 0, \; [\Delta, d] = [\Delta, \delta] 
= 0, \;\Delta = \{ d, \delta\} = (d + \delta)^2$ .} 
$$
\begin{array}{lcl}
|\Psi>_{n} = |\omega>_{n} + \;Q_{b} |\xi>_{n - 1} + \;Q_{d} |\chi>_{n+1}
\;\equiv\;
|\omega>_{n} + \;Q_{ad} |\xi>_{n - 1} +\; Q_{ab} |\chi>_{n+1}.
\end{array} \eqno(1.1)
$$
The above equation encodes the fact that, in the BRST formalism,
there are two sets of conserved charges (viz.$(Q_{b}, Q_{d}, Q_{w})$ and
($Q_{ad}, Q_{ab}, Q_{w})$) that are analogous to the de Rham cohomology
operators of differential geometry. The origin for the existence of this
mapping: $ (Q_{b}, Q_{ad}) \Leftrightarrow d, (Q_{d}, Q_{ab}) \Leftrightarrow
\delta, Q_{w} = \{ Q_{(a)b}, Q_{(a)d} \} \Leftrightarrow \Delta$ has not 
been made 
clear in our earlier works [1-6]. {\it In the present paper, we clarify the 
existence of such a mapping}.

In the physical four $(3 + 1)$-dimensional (4D) flat Minkowskian 
spacetime, the above type of symmetries and corresponding
local and conserved charges have been exploited for the 
definition of the celebrated HDT in the context of free Abelian two-form 
gauge theory [11]. Furthermore, these local and conserved charges 
have been shown to obey exactly the same kind of algebra as that of 
the de Rham cohomology operators of differential geometry defined on
a compact manifold [1,5,6,11]. The existence of
some subtle discrete symmetries has been shown to correspond to the Hodge $*$
duality operation of the ordinary differential geometry for the case of 2D 
one-form-  and 4D two-form Abelian gauge theories [6,11]. Under
these discrete symmetries, the (anti-)BRST- and (anti-)co-BRST symmetry 
transformations are related to one-another in exactly the same way as the 
cohomological operators $d$ and $\delta = \pm * d *$ are related to 
each-other [6]. In our present paper, we show that a proper generalization 
of these discrete symmetries
for the case of a 2D free $U(1)$ gauge theory in the superfield formulation 
(together with the idea of Hodge decomposition  
for the 2D fermionic vector fields) provides a logical origin for the existence of the ``off-shell'' nilpotent, local, covariant
and continuous (anti-)BRST- and (anti-)co-BRST symmetries. The ``on-shell''
version of these results has already been obtained in Ref. [12]. We lay
emphasis on the fact that,
in our present attempt, we concentrate {\it only} on the ``off-shell'' 
version of these nilpotent symmetries which are more general than their
``on-shell'' counterparts. It is evident that we do not exploit the 
classical equations of 
motion {\it anywhere} in our present discussion for the off-shell nilpotent
(anti-)BRST- and (anti-)co-BRST symmetries.

One of the most popular geometrical approaches to the BRST formalism (in the 
language of the Maurer-Cartan equation
and the translation generators) is the superfield formulation [13-18].
In this superfield formulation, so far, only the mathematical power of the 
super exterior derivative $\tilde d$, together with the idea of the
horizontality condition 
\footnote{ This condition is referred to as the ``soul flatness'' in Ref. [18]
implying the flatness of the curvature (two-form) tensor
in the Grassmannian directions of the supermanifold.}
on the two-form constructed with it, has been tapped which leads to the 
interpretation of the nilpotent (anti-)BRST charges as the generators for 
translations along the Grassmannian
directions of the supermanifold for the one-form gauge theory [13-15].
In view of the generality and applicability
of this formulation to any arbitrary $n$-form ($n = 1, 2....$) 
gauge theories in any arbitrary spacetime dimension [15-18], it is an 
interesting endeavour to provide the geometrical interpretation for the 
nilpotent
(anti-)dual BRST symmetries and a bosonic symmetry (generated by the 
Casimir operator) [1-6,11] in the language of other super de Rham cohomology
operators $ \tilde \delta = \pm\; \star  \tilde d \star $ and  $
\tilde \Delta = \tilde d \tilde \delta + \tilde \delta \tilde d$ defined
on the four ($2+2$)-dimensional compact supermanifold.
Here $\tilde \delta$ and $\tilde \Delta$ are the super co-exterior
derivative and super Laplacian operator, respectively. 
To the best of our knowledge, these super cohomological operators have not yet
been exploited in the context of superfield approach to BRST 
formalism, BRST cohomology and HDT in the quantum Hilbert space 
of states for any gauge
theory in any spacetime dimension. 
{\it One of the central themes of the present paper is to make an 
attempt towards this goal and to show that useful symmetries do emerge for 
the 2D free Abelian gauge theory
when we exploit these geometrical (cohomological) super operators
together with the idea of the generalized version of horizontality condition.}
We demonstrate that the off-shell
nilpotent (anti-)dual BRST symmetries (generated by (anti-)co-BRST charges)
emerge when we exploit the super co-exterior derivative 
$\tilde \delta = \pm \;\star \tilde d \star$ 
together with an analogue of the  ``horizontality'' condition.
Geometrically, the nilpotent and conserved (anti-)co-BRST charges
(similar to their counterpart (anti-)BRST charges [15])
turn out to be the translation generators along
the Grassmannian directions 
\footnote{ Even though (anti-)BRST- and (anti-)co-BRST charges generate 
translations
along the Grassmannian directions, there is a clear distinction between them.
To be precise, these charges generate translations for the 
fermionic (anti-)ghost fields along {\it different} Grassmannian 
directions of the supermanifold (cf.Sec.4).} 
of the supermanifold. 
When we exploit the mathematical
power of super Laplacian operator $\tilde \Delta = \tilde d \tilde \delta 
+ \tilde \delta \tilde d$ in this context, we obtain a bosonic symmetry
transformation for the $U(1)$ gauge field {\it alone} and ghost fields do not
transform at all. As expected, the Noether conserved charge corresponding to 
this symmetry emerges as the generator for the 
above symmetry transformations. In the language of the
geometry on the supermanifold, this bosonic charge (Casimir operator)
turns out to be the generator of translations along 
the $\theta\bar\theta$-direction of the supermanifold. In fact, this bosonic 
direction is  equivalent to a couple of successive translations
along the Grassmannian ($\theta$ and $\bar\theta$) directions
of the supermanifold. 

This paper is organized as follows. In section 2, we set up the notations
and recapitulate the bare essentials of our earlier works [1-6] for the 2D 
free Abelian gauge theory in the framework of Lagrangian formulation. This is
followed by the derivation of the (anti-)BRST symmetries in the framework of 
superfield formalism in section 3 by exploiting the idea 
of ``horizontality'' condition [15]. In section 4, we derive the 
(anti-)dual BRST symmetries by exploiting the 
super dual exterior derivative $\tilde \delta$ together with 
a restriction ($ \tilde \delta \tilde A = \delta A$) which
is the analogue of the horizontality condition. Section 5 is 
devoted to the discussion of some new discrete symmetries that are present in
the superfield formulation. In section 6, we derive a bosonic symmetry by
exploiting the mathematical power of $\tilde \Delta$ together with 
a ``horizontality type'' restriction w.r.t. this operator. Finally, 
in section 7, we make some concluding remarks and point out some future 
directions that can be pursued later.\\

\noindent
{\bf 2 Preliminary:  (anti-)BRST- and (anti-)co-BRST symmetries}\\

\noindent
We begin with the BRST invariant Lagrangian density (${\cal L}_{b}$) for the
flat Minkowskian two $(1+1)$-dimensional
\footnote{ We adopt here the notations and conventions
in which the 2D flat Minkowski
metric is : $\eta_{\mu\nu} = $ diag $ (+1, -1) $ and $ \Box = \eta^{\mu\nu}
\partial_{\mu} \partial_{\nu} = \partial_{0} \partial_{0} -
\partial_{1} \partial_{1},\;  F_{01} = - \varepsilon^{\mu\nu} \partial_\mu A_\nu
= \partial_{0} A_{1} - \partial_{1} A_{0} = E
= F^{10}, \;\varepsilon_{01} = \varepsilon^{10} = +1.$}
free Abelian gauge theory in the Feynman gauge (see,  e.g., [18-20])
$$
\begin{array}{lcl}
{\cal L}_{b} = - \frac{1}{4} F^{\mu\nu} F_{\mu\nu}
- \frac{1}{2} (\partial \cdot A)^2 - i \partial_{\mu}\bar C \partial^{\mu} C
\;\equiv\; \frac{1}{2} E^2
- \frac{1}{2} (\partial \cdot A)^2 - i \partial_{\mu}\bar C \partial^{\mu} C
\end{array} \eqno(2.1)
$$
where $F_{\mu\nu} = \partial_{\mu} A_{\nu} - \partial_{\nu} A_{\mu}$
(with $F_{01} = E$ as electric field) is the field strength tensor 
derived from the one-form $A = dx^\mu A_{\mu} $ 
(with $ A_{\mu}$ as the vector potential)
by application of the exterior derivative $d$ (i.e. $F = d A$). On the
other hand, the gauge-fixing term $(\partial \cdot A)$ is derived from the
one-form $ A = dx^\mu A_{\mu}$ by application of the co-exterior derivative
$\delta$ (i.e. $(\partial \cdot A) = \delta A, \delta = - * d * $). Thus, 
two-form $ F = d A$ and zero-form $(\partial \cdot A) = \delta A$ are 
`Hodge' dual to each-other. The (anti-)ghost fields $(\bar C) C$ are
anti-commuting $( \bar C^2 = C^2 = 0, C \bar C = - \bar C C)$ in nature. One
can linearize the kinetic energy term and the gauge-fixing term by introducing
auxiliary-fields ${\cal B}$ and $B$, as 
$$
\begin{array}{lcl}
{\cal L}_{B} = {\cal B}\; E - \frac{1} {2}\;{\cal B}^2 +
B \; (\partial \cdot A) + \frac{1}{2} B^2 - i \partial_{\mu} \bar C
\partial^{\mu} C.
\end{array}\eqno(2.2)
$$
This Lagrangian density respects the following off-shell nilpotent
$(s_{b}^2 = 0, s_{d}^2 = 0)$ BRST ($s_{b}$) and dual-BRST ($s_{d}$) 
symmetry transformations
\footnote{ We follow here the notations adopted in Ref. [21]. In fact, 
a BRST transformation $\delta_{B}$ is the product of an 
anti-commuting constant parameter $\zeta$ and the transformation $s_{b}$ 
(i.e. $\delta_{B} = \zeta\; s_{b}$) in its full glory.}
$$
\begin{array}{lcl}
s_{b} A_{\mu} &=&  \partial_{\mu} C \qquad s_{b} C = 0
\quad s_{b} {\cal B} = 0
\quad s_{b} E = 0 \nonumber\\
s_{b} \bar C &=&  i B \qquad \;s_{b} B = 0 
\qquad \;s_{b} (\partial \cdot A) =  \Box C
\end{array} \eqno(2.3)
$$
$$
\begin{array}{lcl}
s_{d} A_{\mu} &=& -  \varepsilon_{\mu\nu}\; \partial^{\nu} \bar C
\qquad \;s_{d} C = - i {\cal B} \qquad\; s_{d} {\cal B} = 0 \nonumber\\
s_{d} E &=&  \Box \bar C \quad 
s_{d} \bar C = 0 \quad s_{d} B = 0 \quad
s_{d} (\partial \cdot A) = 0. 
\end{array}\eqno(2.4)
$$
The corresponding off-shell nilpotent ($ s_{ab}^2 = 0, s_{ad}^2 = 0$)
anti-BRST ($s_{ab}$) symmetries (with $ s_{b} s_{ab} + s_{ab} s_{b} = 0$)
and anti-co-BRST ($s_{ad}$) symmetries (with $s_{d} s_{ad} 
+ s_{ad} s_{d} = 0$) are
$$
\begin{array}{lcl}
s_{ab} A_{\mu} &=&  \partial_{\mu} \bar C \quad s_{ab} \bar C = 0
\quad s_{ab} {\cal B} = 0
\quad s_{ab} E = 0 \nonumber\\
s_{ab}  C &=& - i B \qquad s_{ab} B = 0 \qquad
s_{ab} (\partial \cdot A) =  \Box \bar C
\end{array} \eqno(2.5)
$$
$$
\begin{array}{lcl}
s_{ad} A_{\mu} &=& - \; \varepsilon_{\mu\nu}\; \partial^{\nu}  C
\qquad \;\;s_{ad} \bar C =  i {\cal B} \qquad \;s_{ad} {\cal B} = 0 \nonumber\\
s_{ad} E &=&  \Box  C \;\quad 
s_{ad} C = 0 \;\quad s_{ad} B = 0 \;\quad
s_{ad} (\partial \cdot A) = 0. 
\end{array}\eqno(2.6)
$$
The anti-commutators of the above symmetries lead to the definition of a 
bosonic symmetry $s_{w} = \{ s_{b}, s_{d} \} 
= \{ s_{ab}, s_{ad} \}, s_{w}^2 \neq 0$. Under
this symmetry, the (anti-)ghost fields do not transform and only the $U(1)$
gauge field transform as [1,2,6]
$$
\begin{array}{lcl}
s_{w} A_{\mu} &=& \; (\partial_{\mu} {\cal B}
+ \varepsilon_{\mu\nu} \;\partial^{\nu} B) \equiv \;\varepsilon_{\mu\nu}\;
(\partial^\nu B + \varepsilon^{\nu\lambda}\; \partial_{\lambda} {\cal B})
\qquad\;\;\; s_{w} B = 0 \nonumber\\
s_{w} C &=& 0 \quad s_{w} \bar C = 0 \quad
s_{w} {\cal B} = 0 \quad
s_{w} (\partial \cdot A) =  \Box {\cal B} \quad
s_{w} E = - \Box B.
\end{array} \eqno (2.7)
$$
It can be seen that the transformation for the $U(1)$ gauge field $A_{\mu}$
is just equal to its own equation of motion: $ \partial_{\mu} {\cal B} 
+ \varepsilon_{\mu\nu} \partial^\nu  B (= 0$).  
The substitutions $ {\cal B} = E, B = - (\partial \cdot A)$ make
the above equation of motion equal to $-\varepsilon_{\mu\nu} \Box A^\nu (= 0)$.
In terms of the vector fields $A_\mu$ alone, now we have the transformation
$\tilde s_{w} A_\mu = - \varepsilon_{\mu\nu} \Box A^\nu$. Thus, we note that 
all the transformations $s_{w}$ in (2.7) (and their analogues $\tilde s_{w}$)
are trivially zero on the on-shell 
\footnote{ We shall be exploiting this observation in Secs. 6 and 7.
It will be noticed, however, that 
$\tilde s_{w}$ (with $\tilde s_{w} C = 0, \tilde s_{w} \bar C = 0,
\tilde s_{w} A_\mu = -\varepsilon_{\mu\nu} \Box A^\nu$) is a symmetry 
transformation for the Lagrangian density (2.1) and is the analogue of 
the symmetry $s_{w}$ (cf.(2.7)) 
for the Lagrangian density (2.2).}.  All the above 
continuous symmetry transformations can be concisely expressed, in terms of 
generators $Q_{r}$ (see, e.g., Sec. 7 for explicit local expressions)
and generic field $\phi$,  as
$$
\begin{array}{lcl}
s_{r} \phi = - i\; [\; \phi, Q_{r} \;]_{\pm}\; \qquad \;r = b, ab, d, ad, w
\end{array} \eqno(2.8)
$$
where $Q_{r}$ are the conserved Noether charges derived from the conserved
currents and brackets $[\; ,\; ]_{\pm}$
stand for the (anti-)commutators for $\phi$ being (fermionic)bosonic in nature.

Now we shall dwell a bit on the presence of some 
interesting discrete symmetries in the
theory. It can be seen from (2.3) and (2.5) that $s_{b} \leftrightarrow s_{ab}$
for the discrete transformations: $C \leftrightarrow \bar C, B \leftrightarrow
- B$. In a similar fashion, it is clear from (2.4) and (2.6) that $s_{d}
\leftrightarrow s_{ad}$ for the presence of the discrete transformations:
$ C \leftrightarrow \bar C, {\cal B} \leftrightarrow - {\cal B}$. In other
words, the discrete symmetries: $ C \leftrightarrow \bar C, B \leftrightarrow
- B, {\cal B} \leftrightarrow - {\cal B}, A_{\mu} \leftrightarrow A_{\mu}$,
connect the BRST to anti-BRST and the co-BRST to anti-co-BRST symmetries and 
vice-versa. Yet another interesting discrete symmetry transformations [1,2,6]
$$
\begin{array}{lcl}
&&C \rightarrow \pm i \bar C \qquad  E \rightarrow \pm i (\partial \cdot A)
\qquad A_{\mu} \rightarrow A_{\mu} \qquad \partial_{\mu} \rightarrow 
\pm i \varepsilon_{\mu\nu} \partial^\nu  \nonumber\\
&&\bar C \rightarrow \pm i C \qquad (\partial \cdot A)
\rightarrow \pm i E \qquad {\cal B} \rightarrow \mp i\; B
\qquad B \rightarrow \mp i\; {\cal B}  
\end{array}\eqno(2.9)
$$
connect $s_{b}$ with $s_{d}$ (as well as $s_{ab}$ with $s_{ad}$) and
{\it leave the Lagrangian density (2.2) form-invariant} [6]. 
The topological nature of this theory
has been shown by demonstrating that the Lagrangian density in (2.1) can
be written, modulo some total derivatives, as the sum of BRST- and dual BRST
anti-commutators [1,2,6]
$$
\begin{array}{lcl}
{\cal L}_{b} &=& \{ Q_{d}, T_{1} \} + \{ Q_{b}, T_{2} \} \equiv
\;s_{d}\; ( i T_{1} )\; + \;s_{b}\; (i T_{2})\nonumber\\
{\cal L}_{b} &=& \{ Q_{ad}, P_{1} \} + \{ Q_{ab}, P_{2} \} \equiv
\;s_{ad}\; (i P_{1})\; + \;s_{ab}\; (i P_{2})\nonumber\\
\end{array} \eqno(2.10)
$$
where the local expressions for $T_{(1,2)}$ and $P_{(1,2)}$ are:
$ T_{1} = \frac{1}{2} E C, T_{2} =- \frac{1}{2} (\partial \cdot A) \bar C,
 P_{1} = - \frac{1}{2} E \bar C, P_{2} = \frac{1}{2} (\partial \cdot A) C$.
Similarly, the off-shell nilpotent (anti-)BRST- and (anti-)dual BRST
invariant Lagrangian density (2.2) can be written in two different ways as:
(i) the sum of the kinetic energy term and an (anti-)BRST invariant part
or (ii) the sum of the gauge-fixing term and an (anti-)dual BRST invariant 
part. These alternative ways of expressing the Lagrangian density are
$$
\begin{array}{lcl}
{\cal L}_{B} = {\cal B} E - \frac{1}{2} {\cal B}^2  + s_{b} s_{ab}
( \frac{i}{2} A^2 - \frac{1}{2} \bar C C) 
\equiv B(\partial \cdot A) + \frac{1}{2} B^2
+ s_{d} s_{ad} (\frac{i}{2} A^2 - \frac{1}{2} \bar C C).
\end{array} \eqno(2.11)
$$
The topological invariants for this theory and their well-known recursion
relations have been computed on a 2D compact manifold [1,2,6]. 
Energy-momentum tensor
has been shown to be the sum of BRST- and co-BRST anti-commutators [2,6].\\

\noindent             
{\bf 3  Superfield formulation for (anti-)BRST symmetry transformations}\\

\noindent
We start off with a four ($ 2 + 2$) dimensional supermanifold
that is parametrized by two c-number commuting (bosonic) spacetime
coordinates $ x^\mu \;(\mu = 0, 1) $ and two Grassmann variables $\theta$
and $\bar \theta$. We define a supervector superfield $v_{s}$ on this
supermanifold as [15]
\footnote {We follow the field notations of Ref. [13,15,17] but be consistent
with the conventions adopted in Ref. [22] for the definition of the super 
derivatives and super differential forms on a compact supermanifold.}
$$
\begin{array}{lcl}
v_{s} = \;\Bigl ( \Phi_{\mu}\; (x, \theta, \bar \theta), \eta\; (x, \theta, 
\bar \theta), \bar \eta \;(x, \theta, \bar \theta) \Bigr )
\end{array}\eqno(3.1)
$$
where $\Phi_{\mu} (x, \theta, \bar \theta)$ are the bosonic (even)
superfields and $\eta (x, \theta, \bar \theta), 
\bar \eta (x, \theta, \bar \theta)$ are the fermionic (odd) superfields
which constitute the supermultiplet of the supervector superfield $v_{s}$.
In terms of the superspace coordinates $x^\mu, \theta, \bar \theta$, the
component superfields can be expanded as
$$
\begin{array}{lcl}
\Phi_{\mu}\; (x, \theta, \bar \theta)
&=& A_{\mu} (x) + \theta \;\bar R_{\mu} (x) + \bar \theta\; R_{\mu} (x)
+ i \;\theta \;\bar \theta \;S_{\mu} (x) \nonumber\\
\eta \;(x, \theta, \bar \theta)
&=& C (x) + i \;\theta \;\bar B (x) + i \;\bar \theta \;{\cal B} (x)
+ i \;\theta \;\bar \theta \;s (x)\nonumber\\
\bar \eta \;(x, \theta, \bar \theta)
&=& \bar C (x) + i \;\theta \;\bar {\cal B} (x) + i \;\bar \theta\;  B (x)
+ i \;\theta \;\bar \theta \;\bar s (x)
\end{array}\eqno(3.2)
$$
where field variables $A_{\mu} (x) \equiv A_{\mu} (x, 0, 0); 
C (x) \equiv C (x, 0, 0), etc.$ are functions of only spacetime variables.
It will be noticed that the local 
fields: $A_{\mu}(x), S_{\mu}(x), B(x), \bar B(x), 
{\cal B}(x), \bar {\cal B}(x)$ are bosonic (even) in nature whereas 
the local fermionic (odd) fields in the theory are: $ R_{\mu}(x), 
\bar R_{\mu}(x), C(x), \bar C(x), s(x), \bar s(x)$. 
The super exterior derivative $\tilde d = d z^M \partial_{M}$ on this 
supermanifold is defined in terms of the superspace differentials 
($dx^\mu, d \theta, d \bar\theta$) as [22]
$$
\begin{array}{lcl}
\tilde d = dx^\mu \;\partial_{\mu} + d \theta\; \partial_{\theta}
+ d \bar \theta\; \partial_{\bar \theta}
\end{array}\eqno(3.3)
$$
where
$$
\begin{array}{lcl}
&&\tilde d = d z^M {\displaystyle \frac{\partial}{\partial z^M}} \qquad
\tilde d^2 = 0 \quad z^M = (x^\mu, \theta, \bar \theta) \nonumber\\
&&\partial_{\mu} = {\displaystyle \frac{\partial} {\partial x^\mu}} \qquad
\partial_{\theta} = {\displaystyle \frac{\partial}{\partial \theta}} \qquad
\partial_{\bar \theta} = {\displaystyle \frac{\partial}{\partial \bar \theta}}.
\end{array} \eqno(3.4)
$$
A connection super one-form on the supermanifold can be defined in terms of the
component superfields of the supervector superfield $v_{s}$ as
$$
\begin{array}{lcl}
\tilde A = d x^\mu \;\Phi_{\mu} (x, \theta, \bar \theta)
+ d \theta\; \bar \eta (x, \theta, \bar \theta)
+ d \bar \theta \; \eta (x, \theta, \bar \theta).
\end{array}\eqno(3.5)
$$
The Maurer-Cartan equation that defines the curvature two-form $\tilde F$
from the connection one-form $\tilde A$ and the exterior derivative $\tilde d$
on the supermanifold, in its most general form,  is
$$
\begin{array}{lcl}
\tilde F = \tilde d \tilde A + \tilde A \wedge \tilde A
\equiv \tilde d \tilde A + \frac{1}{2}\; [ \tilde A, \tilde A ]. 
\end{array} \eqno(3.6)
$$
For Abelian $U(1)$ gauge theory, however, the last term in the above equation
is zero (i.e., $ \tilde A \wedge \tilde A 
= \frac{1}{2} [ \tilde A, \tilde A ] = 0)$. The horizontality condition imposes 
the restriction that the components of the curvature two-form 
$\tilde F= \tilde d \tilde A$ for the Abelian gauge theory (with
$\{ d \theta (d \bar \theta), \theta (\bar \theta \} = 0$)
$$
\begin{array}{lcl}
\tilde F = \tilde d \tilde A &=& (d x^\mu \wedge d x^\nu) \;(\partial_{\mu} 
\Phi_{\nu}) + \;(d x^\mu \wedge d \theta) \; (\partial_{\mu} \bar \eta 
- \partial_{\theta} \Phi_{\mu}) -  \;(d \theta \wedge d \theta) 
\; (\partial_{\theta} \bar \eta)\nonumber\\
&+& (d x^\mu \wedge d \bar \theta) \;(\partial_\mu \eta - \partial_{\bar \theta}
\Phi_{\mu}) - (d \theta \wedge  d \bar \theta) \; (\partial_{\theta} \bar \eta
+ \partial_{\bar \theta} \eta) - (d \bar \theta \wedge d \bar \theta) \;
(\partial_{\bar \theta} \eta)
\end{array} 
$$
must vanish along the Grassmannian directions ($\theta , \bar \theta$) of the
supermanifold. This ultimately amounts to the following condition (which is
nothing but $\tilde d \tilde A = d A$)
$$
\begin{array}{lcl}
\tilde F \equiv \frac{1}{2} \;(d z^M \wedge d z^N) \; \tilde F_{MN} 
= \frac{1}{2}\;(dx^\mu \wedge d x^\nu) \; F_{\mu\nu} \equiv F
\end{array} \eqno(3.7)\\
$$
where the wedge product $d z^M \wedge d z^N$, in the component form of
the superspace variables, is: $ d x^\mu \wedge d x^\nu = - d x^\nu \wedge
d x^\mu, \; d \theta \wedge d \bar \theta = d \bar \theta \wedge d \theta,
d x^\mu \wedge d \theta = - d \theta \wedge d x^\mu,$ etc.
In the following, the above horizontality condition (3.7) leads to
$$
\begin{array}{lcl}
\partial_{\theta} \bar \eta = 0 &\rightarrow &\bar {\cal B} = 0 
\qquad \bar s = 0 \nonumber\\
\partial_{\bar \theta}  \eta = 0 &\rightarrow&  {\cal B} = 0 
\qquad  s = 0 \nonumber\\
 \partial_{\theta} \eta = - \partial_{\bar \theta} \bar \eta 
&\rightarrow& \; B + \bar B = 0 \nonumber\\
 \partial_{\mu} \eta = \partial_{\bar \theta} \Phi_{\mu} &\rightarrow&
R_{\mu} = \partial_{\mu} C \qquad S_{\mu} = - \partial_{\mu} \bar B
\equiv \partial_{\mu} B \nonumber\\
\partial_{\mu} \bar \eta = \partial_{ \theta} \Phi_{\mu} &\rightarrow&
\bar R_{\mu} = \partial_{\mu} \bar C \qquad S_{\mu} = 
\partial_{\mu} B 
\end{array}\eqno(3.8)
$$
which satisfy the other conditions, namely;
$$
\begin{array}{lcl}
\partial_{\mu} \bar R_{\nu} - \partial_{\nu} \bar R_{\mu} = 0 \qquad
\partial_{\mu}  R_{\nu} - \partial_{\nu}  R_{\mu} = 0 \qquad
\partial_{\mu}  S_{\nu} - \partial_{\nu}  S_{\mu} = 0.
\end{array}\eqno(3.9)
$$
It is clear (cf.Sec.2) that the (anti-)BRST charges are the generators 
(cf.(2.8)) of translations along the Grassmannian directions 
(i.e., $ \mbox{Lim}_{\theta,\bar\theta \rightarrow 0}
\; \frac{\partial}{\partial \theta (\bar\theta)}\;
\Sigma (x,\theta,\bar\theta) = i [Q_{(a)b}, \Lambda]_{\pm}$, for $\Sigma
= \eta, \bar \eta, \Phi_\mu$ and corresponding
$\Lambda = C, \bar C, A_\mu$) of the 
supermanifold as the expansions in (3.2) can be recast in terms of
$s_{(a)b}$ as
$$
\begin{array}{lcl}
\Phi_{\mu}\; (x, \theta, \bar \theta)
&=& A_{\mu} (x) + \theta \;(s_{ab} A_{\mu}(x)) 
+ \bar \theta \;(s_{b} A_{\mu}(x))
+  \theta \;\bar \theta \;(s_{b} s_{ab} A_{\mu}(x)) \nonumber\\
\eta \;(x, \theta, \bar \theta)
&=& C (x) + \theta \;(s_{ab} C(x)) + \bar \theta \;(s_{b} C(x))
+ \theta \;\bar \theta \;(s_{b} s_{ab} C(x)) \nonumber\\
\bar \eta \;(x, \theta, \bar \theta)
&=& \bar C (x) + \theta \;(s_{ab} \bar C(x)) 
+  \bar \theta \; (s_{b} \bar C(x))
+ \theta\; \bar \theta \;(s_{b} s_{ab} \bar C(x))
\end{array}\eqno(3.10)
$$
which finally amounts to the following expansion
$$
\begin{array}{lcl}
\Phi_{\mu} (x, \theta, \bar \theta)
&=& A_{\mu} (x) + \theta \;\partial_{\mu} \bar C (x)
+ \bar \theta \;\partial_{\mu} C (x)
+ i \;\theta \;\bar \theta \;\partial_{\mu} B (x) \nonumber\\
\eta (x, \theta, \bar \theta)
&=& C (x) + i \;\theta \; \bar B (x) \equiv
 C (x) - i \;\theta \; B (x) \nonumber\\
\bar \eta (x, \theta, \bar \theta)
&=& \bar C (x) + i \;\bar \theta \; B (x).
\end{array}\eqno(3.11)
$$
This, in a nut-shell, provides the origin for the existence of (anti-)BRST
symmetries in the framework of superfield formulation. \\

\noindent
{\bf 4 Superfield approach to (anti-)dual BRST transformations}\\

\noindent
On an even dimensional {\it ordinary} spacetime manifold, it is obvious that
the operation of $ \delta = - * d * $ 
\footnote{ In any arbitrary flat Minowskian even $D$-dimensional ordinary
spacetime manifolds, the ordinary dual-exterior derivative $\delta$
is: $\delta = - * d * $. In general, an inner-product of a $n$-form
in $D$-dimensional spacetime manifold leads to $\delta = (-1)^{D n + D +1}
* d *$. Thus, for odd $D$, we have $\delta = (-1)^n * d *$ (see, e.g. 
Ref. [7]).}
on the one-form $ A = dx^\mu A_{\mu}$
leads to the zero-form as none other than the  gauge-fixing term 
$(\partial \cdot A)$ [7,8]: 
$$
\begin{array}{lcl}
\delta A = - * d * A = (\partial \cdot A)
\end{array}\eqno(4.1)
$$
where the Hodge $*$ duality operation on the ordinary differentials
in the flat Minkowskian two-dimensional spacetime is:
$ * \;(dx^\mu) = \varepsilon^{\mu\nu}\; (dx_{\nu}), \;
* \;(dx^\mu \wedge dx^\nu) = \varepsilon^{\mu\nu}.$ Now, on the connection 
super one-form  $\tilde A$ defined on a $2 +2$-dimensional supermanifold,
we apply the super co-exterior derivative $\tilde \delta$ and exploit the
following condition 
$$
\begin{array}{lcl}
\tilde \delta \tilde A = - \star \;\tilde d 
\;\star \;\tilde A = (\partial \cdot A).
\end{array}\eqno (4.2)
$$
In words, this equation amounts to restricting the zero-form superscalar 
superfield ($\tilde \delta \tilde A$), that emerges after the application 
of $\tilde \delta$ on the connection super one-form $\tilde A$, to the 
zero-form gauge-fixing term ($\delta A = (\partial \cdot A)$)
defined on the ordinary flat 2D Minkowskian spacetime manifold.
This  restriction is the analogue of horizontality condition w.r.t.
$\tilde \delta$. Here the
Hodge $\star$ duality operation is defined  on the 
$(2+2)$-dimensional compact supermanifold. The basic 
superspace differentials transform under this operation as:
$$
\begin{array}{lcl}
&& \star\;(d x^\mu)\; = \;\varepsilon^{\mu\nu}\; (dx_{\nu})\; \;\qquad \;\;
\star \;(d \theta)  = \; (d \bar \theta) \;\;\qquad\;\;
\star \;(d \bar \theta) = \;(d \theta) \;\; 
\nonumber\\ && \star\; (d\theta \wedge d \theta)
= \;s^{\theta \theta} \;\; \qquad \; \star \;(d \bar \theta 
\wedge d \bar \theta) = \;
s^{\bar \theta \bar \theta} \;\;\qquad\;\;
\star\; (d \theta \wedge d \bar \theta) 
= \;s^{\theta \bar \theta}\; \nonumber\\
&& \star\; (d x^\mu \wedge d x^\nu) = \;\varepsilon^{\mu\nu} \;\qquad
\star\; (d x^\mu \wedge d \theta) = \varepsilon^{\mu\theta} \;\qquad
\star\; (d x^\mu \wedge d \bar \theta) = \varepsilon^{\mu \bar \theta}
\end{array}\eqno (4.3)
$$
where $\varepsilon^{\mu \theta (\bar \theta)} = - \varepsilon^{\theta 
(\bar \theta)\mu}$ are anti-symmetric and $ s^{\theta \bar \theta} 
= s^{\bar \theta \theta}$ etc. are symmetric.
The expansion of the l.h.s. of (4.2) defines a 
superscalar (zero-form) superfield
$$
\begin{array}{lcl}
\tilde \delta \tilde A &=& (\partial \cdot \Phi) + s^{\theta\theta}\;
(\partial_{\theta} \eta) + s^{\bar\theta \bar\theta}\; (\partial_{\bar \theta}
\bar \eta) + s^{\theta \bar \theta} \;(\partial_{\theta} \bar \eta +
\partial_{\bar \theta} \eta) \nonumber\\
&-& \varepsilon^{\mu\theta} \;(\partial_\mu \eta + \varepsilon_{\mu\nu}
\partial_{\theta} \Phi^{\nu}) - \varepsilon^{\mu  \bar \theta}\;
(\partial_\mu \bar \eta + \varepsilon_{\mu\nu} \partial_{\bar \theta}
\Phi^{\nu})
\end{array}
$$
and its subsequent equality with the ordinary gauge-fixing term in the 
r.h.s. (due to the analogue of the horizontality requirement) leads to 
$$
\begin{array}{lcl}
\partial_{\bar \theta} \bar \eta = 0 &\rightarrow & B = 0
\qquad \bar s = 0 \nonumber\\
\partial_{\theta}  \eta = 0 &\rightarrow&  \bar B = 0 
\qquad  s = 0 \nonumber\\
 \partial_{\theta} \bar \eta = - \partial_{\bar \theta} \eta 
&\rightarrow& \; {\cal B} + \bar {\cal B} = 0 \nonumber\\
 \partial_{\mu} \; \bar \eta 
= - \varepsilon_{\mu\nu} \;\partial_{\bar \theta}\; \Phi^\nu &\rightarrow&
R_{\mu} = - \varepsilon_{\mu\nu} \;\partial^{\nu} \bar C 
\qquad S_{\mu} = + \varepsilon_{\mu\nu}\;\partial^{\nu} \bar {\cal B}
\equiv - \varepsilon_{\mu\nu} \;\partial^{\nu} {\cal B} 
\nonumber\\
\partial_{\mu} \; \eta = - \varepsilon_{\mu\nu}
\partial_{ \theta} \Phi^{\nu} &\rightarrow&
\bar R_{\mu} = - \varepsilon_{\mu\nu}\;
\partial^{\nu} C \qquad S_{\mu} = -\varepsilon_{\mu\nu}\;\partial^\nu {\cal B}. 
\end{array}\eqno(4.4)
$$
The  above results also satisfy the following conditions:
$$
\begin{array}{lcl}
\partial \cdot \bar R = 0 \;\qquad \partial \cdot R = 0 \;\qquad
\partial \cdot S = 0.
\end{array}\eqno(4.5)
$$
It will be noticed that {\it (4.4) can be directly obtained from (3.8) 
by exploiting the duality symmetry (2.9)}.  In the language of the expansion 
of the superfields, we have obtained 
$$
\begin{array}{lcl}
\Phi_{\mu} (x, \theta, \bar \theta)
&=& A_{\mu} (x) - \theta \;\varepsilon_{\mu\nu} \partial^{\nu}  C (x)
- \bar \theta \;\varepsilon_{\mu\nu} \partial^{\nu} \bar C (x)
- i \theta \;\bar \theta \;\varepsilon_{\mu\nu}
\partial^{\nu} {\cal B} (x)\nonumber\\
\eta (x, \theta, \bar \theta)
&=& C (x) + i \;\bar \theta \; {\cal B} (x)\nonumber\\
\bar \eta (x, \theta, \bar \theta)
&=& \bar C (x) + i \; \theta \; \bar {\cal B} (x) \equiv
 \bar C (x) - i \; \theta \;  {\cal B} (x).
\end{array}\eqno (4.6)
$$
This result can be further written in terms of the (anti-)dual BRST 
$s_{(a)d}$ symmetries (cf. eqns. (2.4) and (2.6))
obtained in Sec. 2, as follows
$$
\begin{array}{lcl}
\Phi_{\mu}\; (x, \theta, \bar \theta)
&=& A_{\mu} (x) + \theta \;(s_{ad} A_{\mu}(x))
+ \bar \theta \;(s_{d} A_{\mu}(x))
-  \theta \;\bar \theta \;(s_{d} s_{ad} A_{\mu}(x)) \nonumber\\
\eta \;(x, \theta, \bar \theta)
&=& C (x) + \theta \;(s_{ad} C(x)) - \bar \theta \;(s_{d} C(x))
+ \theta \;\bar \theta \;(s_{d} s_{ad} C(x)) \nonumber\\
\bar \eta \;(x, \theta, \bar \theta)
&=& \bar C (x) - \theta \;(s_{ad} \bar C(x))  
+  \bar \theta \; (s_{d} \bar C(x))
+ \theta \;\bar \theta \;(s_{d} s_{ad} \bar C(x)).
\end{array}\eqno(4.7)
$$
These equations (4.5--4.7), finally, establish the origin of the existence
of (anti-)dual BRST symmetries (discussed in Sec.2) in the framework of 
superfield formalism. It is evident that application of a single
restriction (i.e., the analogue of horizontality condition) with 
the super co-exterior
derivative (operating on $\tilde A$) leads to the derivation of two 
nilpotent (anti-) dual BRST symmetries for the 2D free Abelian gauge theory.
It is worthwhile to compare and contrast transformations (3.11) vis-a-vis
transformations (4.6) that are generated by (anti-)BRST- and (anti-)co-BRST
charges $Q_{(a)b}$ and $Q_{(a)d}$, respectively. These are: (i) (anti-)BRST-
and (anti-)co-BRST transformations correspond to translations along $\theta$
and $\bar \theta$ directions of the supermanifold for bosonic- and fermionic
fields of the theory. (ii) For the bosonic field $A_\mu$, the above
nilpotent charges generate translations along $\theta, \bar \theta$ and
$\theta\bar\theta$ directions of the supermanifold. (iii) For the fermionic
fields $C$ and $\bar C$, the  translations are either 
along $\theta$ or $\bar \theta$ directions of the supermanifold. 
(iv) For the (anti-)ghost fields $(\bar C)C$,
the (anti-)BRST charges generate translations along $(\bar \theta)\theta$ 
directions of the supermanifold. On the contrary, for the same fields,
the (anti-)co-BRST charges generate translations along $(\theta)\bar\theta$
directions. (v) It is the mathematical power of $\tilde d$ (together with the
Maurer-Cartan equation) that produces (anti-)BRST transformations. On the
other hand, the derivation of the (anti-)co-BRST transformations depends
entirely on the mathematical power of $\tilde \delta$. (vi) The mapping between
(anti-)BRST- and (anti-)co-BRST charges on the one hand and the super
 cohomological operators $\tilde d$ and $\tilde \delta$ on the other hand,
is: $Q_{(a)b} \Leftrightarrow \tilde d$ and $ Q_{(a)d} \Leftrightarrow \tilde
\delta$. (vii) The local conserved charges $Q_{(a)b}$ and $Q_{(a)d}$
are connected with the {\it ordinary} cohomological operators $d$ and $\delta$
through the mappings: $ (Q_{b}, Q_{ad}) \Leftrightarrow d, 
(Q_{d}, Q_{ab}) \Leftrightarrow \delta$ [1,2,6].\\

\noindent
{\bf 5 Discrete symmetry transformations in superfield formulation}\\

\noindent
We have discussed the existence of a few discrete symmetries in Sec. 2. They
are interesting in the sense that they enable us to gain some
insights into various connections which exist among all BRST type symmetries 
that are present in the theory. In fact, it can be checked that the discrete
transformations: $ C \leftrightarrow \bar C, B \leftrightarrow - B,
-{\cal B} \leftrightarrow - {\cal B}$ (that connect BRST- to anti-BRST and
co-BRST- to anti-co-BRST symmetries) can be generalized vis-a-vis super
expansion (3.2) in terms of superspace coordinates and component fields as
$$
\begin{array}{lcl}
C \leftrightarrow \bar C, \quad B \leftrightarrow - B, \quad
{\cal B} \leftrightarrow - {\cal B}, \quad \theta \leftrightarrow \bar \theta,
\quad R_\mu \leftrightarrow \bar R_\mu, \quad
S_\mu \leftrightarrow - S_\mu,
\end{array} \eqno(5.1)
$$
under which the superfields transform as: $\Phi_\mu (x, \theta, \bar \theta)
\leftrightarrow \Phi_\mu (x, \theta, \bar \theta), \eta (x, \theta, 
\bar \theta) \leftrightarrow \bar \eta (x, \theta, \bar \theta)$. It will
be noticed that for this discussion, we have taken $s (x) = \bar s(x) = 0$
and $ B + \bar B = 0, {\cal B} + \bar {\cal B} = 0$ in expansion (3.2), i.e., 
$$
\begin{array}{lcl}
\Phi_{\mu}\; (x, \theta, \bar \theta)
&=& A_{\mu} (x) + \theta \;\bar R_{\mu} (x) + \bar \theta\; R_{\mu} (x)
+ i \;\theta \;\bar \theta \;S_{\mu} (x) \nonumber\\
\eta \;(x, \theta, \bar \theta)
&=& C (x) - i \;\theta \; B (x) + i \;\bar \theta \;{\cal B} (x)
\nonumber\\
\bar \eta \;(x, \theta, \bar \theta)
&=& \bar C (x) - i \;\theta \; {\cal B} (x) + i \;\bar \theta\;  B (x)
\end{array}
$$
due to the fact that these restrictions emerge when we show the existence of 
(anti-)BRST- and (anti-)co-BRST symmetries in the framework of superfield
formulation.  The Hodge decomposed versions for the 2D
vector fields ($R_\mu, \bar R_\mu , S_\mu$), consistent with transformations
(5.1), can be written in terms of the ghost- and auxiliary fields as
$$
\begin{array}{lcl}
R_\mu = \partial_\mu C - \varepsilon_{\mu\nu} \partial^\nu \bar C \qquad
\bar R_\mu = \partial_\mu \bar C - \varepsilon_{\mu\nu} \partial^\nu  C \qquad
S_\mu = \partial_\mu B - \varepsilon_{\mu\nu} \partial^\nu {\cal B} 
\end{array} \eqno(5.2)
$$
which obey $R_\mu \leftrightarrow \bar R_\mu, S_\mu \leftrightarrow - S_\mu$
under $ C \leftrightarrow \bar C, B \leftrightarrow - B, {\cal B}
\leftrightarrow - {\cal B}$. Now we wish to concentrate on the discrete
symmetry transformations (2.9). This symmetry can be generalized in the
framework of superfield formulation as
$$
\begin{array}{lcl}
C &\rightarrow& \pm i \bar C \qquad B \rightarrow \mp i {\cal B} \qquad
(\partial \cdot A) \rightarrow \pm i E \qquad A_\mu \rightarrow A_\mu
\nonumber\\
\bar C &\rightarrow& \pm i C \qquad {\cal B} \rightarrow \mp i B \qquad
E \rightarrow \pm i (\partial \cdot A)  \qquad S_\mu \rightarrow S_\mu
\nonumber\\
R_\mu &\rightarrow& - R_\mu, \quad  \bar R_\mu \rightarrow - \bar R_\mu,
\quad \theta \rightarrow - \theta  \quad \bar \theta \rightarrow -
\bar \theta \quad \partial_\mu \rightarrow \pm i \varepsilon_{\mu\nu}
\partial^\nu
\end{array} \eqno(5.3)
$$
under which the superfields transform as:
$$
\begin{array}{lcl}
\Phi_\mu (x, \theta, \bar \theta) \rightarrow \Phi_\mu (x, \theta, \bar \theta)
\qquad \eta (x, \theta, \bar \theta) \rightarrow \pm i \bar \eta (x, \theta,
\bar \theta)
\qquad \bar \eta (x, \theta, \bar \theta) \rightarrow \pm i  \eta (x, \theta,
\bar \theta).
\end{array} \eqno(5.4)
$$
In the limit: $\theta \rightarrow 0, \bar \theta
\rightarrow 0$, we get back transformations (2.9) from (5.3). It
is interesting to point out that all the basic properties of the superfields
$ \eta^2 = \bar \eta^2 = 0, \eta \bar \eta + \bar \eta \eta = 0, [\Phi_\mu,
\Phi_\nu ] = 0$ remain unchanged under all the discrete symmetry 
transformations (5.1) and (5.3). Now the Hodge decomposed versions for the
2D vectors ($R_\mu, \bar R_\mu, S_\mu$), consistent with the discrete
transformations (5.3), are
$$
\begin{array}{lcl}
R_\mu = \partial_\mu C + \varepsilon_{\mu\nu} \partial^\nu \bar C \qquad
\bar R_\mu = \partial_\mu \bar C + \varepsilon_{\mu\nu} \partial^\nu  C \qquad
S_\mu = \partial_\mu B + \varepsilon_{\mu\nu} \partial^\nu {\cal B} 
\end{array} \eqno(5.5)
$$
which transform as: $R_\mu \rightarrow - R_\mu, \bar R_\mu \rightarrow - \bar
R_\mu, S_\mu \rightarrow S_\mu$ under the transformations: $ C \rightarrow
\pm i \bar C, \bar C \rightarrow \pm i C, B \rightarrow \mp i {\cal B},
{\cal B} \rightarrow \mp i B, \partial_\mu \rightarrow \pm i 
\varepsilon_{\mu\nu} \partial^\nu$. It should be pointed out here that in
the derivation of the
decompositions (5.2) and (5.5), the continuous symmetry transformations
derived in (3.8), (3.11), (4.4) and (4.6) do play an important role.

It is important to note that the form of the Hodge decomposed versions (5.2) 
differs from the decompositions (5.5) {\it only}
by a sign factor. Thus, the decompositions
in (5.2) and (5.5) are {\it orthogonal} to each-other. This shows that the
requirement of the consistency of the Hodge decompositions
with the discrete symmetries of the theory forces two 
independent BRST-type nilpotent, continuous and covariant symmetries for the 
theory. These symmetries are none other than the (anti-)BRST- and 
the (anti-)co-BRST symmetries, we have discussed so far in the
previous sections. {\it It is very interesting to note that
the Hodge decomposed versions for the
2D fermionic vectors $R_\mu$ and $\bar R_\mu$ in (5.2) are also invariant
under the discrete symmetry (duality) transformations (2.9)}. Thus, they can
be called as ``self-dual'' as they transform to themselves under (2.9).\\

\noindent
{\bf 6 Superfield approach to bosonic symmetry transformation}\\

\noindent
It is useful for our discussions to calculate the operation of the
ordinary Laplacian operator $ \Delta = d \delta + \delta d$ on one-form
$ A = d x^\mu A_{\mu}$ defined on the 2D flat 
ordinary Minkowskian spacetime manifold.
In fact, it turns out that the one-form $\Delta A$ can be expressed as
$$
\begin{array}{lcl}
\Delta \;(d x^\mu A_{\mu}) = d x^\mu \;[\; \partial_{\mu} (\partial \cdot A)
- \varepsilon_{\mu\nu} \partial^\nu E  \;] \equiv d x^\mu \;\Box A_{\mu}.
\end{array} \eqno(6.1)
$$
Thus, it is
obvious that the operation of $\Delta$ on a form does not change the degree
of the form. Equation (6.1) is going
to be useful for the application of the analogue of the horizontality 
condition in the context of the operation of super Laplacian operator on
the connection super one-form $\tilde A$. This condition can be succinctly
expressed as the following equality
$$
\begin{array}{lcl}
\tilde \Delta \tilde A = (\tilde d \tilde \delta 
+ \tilde \delta \tilde d) \;\tilde A = d x^\mu \;\Box A_{\mu}.
\end{array} \eqno(6.2)
$$
In other words, we expand the l.h.s. of the above equation in terms of 
the superspace differentials and set equal to zero all
the components of $\tilde \Delta \tilde A$ that are found to be directed
towards the Grassmannian directions $d \theta$ and $d \bar \theta$. In fact,
(6.2) can be expressed as
$$
\begin{array}{lcl}
\tilde d  \;\tilde \delta \;\tilde A &=&
d x^\rho \; J_{\rho} + d \theta \;J_{\theta}  
+ d \bar \theta\; J_{\bar \theta}\nonumber\\
\tilde \delta \;\tilde d \;\tilde A &= & d x^\rho\; H_{\rho} + d \theta\;
H_{\theta} + d \bar \theta \; H_{\bar \theta} 
\end{array} \eqno(6.3)
$$
where the explicit expressions for the $J's$ are:
$$
\begin{array}{lcl}
J_{\rho} &=& \partial_{\rho} \;(\partial \cdot \Phi) 
- \varepsilon^{\lambda \bar \theta}\; \partial_{\rho}\;
 \bigl (\;\partial_\lambda\;
\bar \eta + \varepsilon_{\lambda \mu}\; \partial_{\bar\theta}\;
 \Phi^{\mu} \bigr )
+ s^{\theta\theta} \partial_{\rho}\; \partial_{\theta}\; \eta \nonumber\\
&+& s^{\bar\theta\bar\theta}\; \partial_{\rho}\; \partial_{\bar\theta}\;
 \bar \eta
- \varepsilon^{\lambda \theta}\; \partial_{\rho}\; \bigl (\;\partial_\lambda
\eta + \varepsilon_{\lambda \mu}\; \partial_{\theta}\; \Phi^{\mu} \bigr )
+ s^{\theta\bar\theta} \;\partial_{\rho}\;\bigl (\;\partial_{\theta} \bar \eta 
+ \partial_{\bar \theta}\; \eta \bigr )\nonumber\\
J_{\theta} &=& s^{\bar\theta\bar\theta}\; \partial_{\theta} \;
\partial_{\bar \theta} \;\bar \eta + \partial_{\theta}\; (\partial \cdot \Phi)
- \varepsilon^{\lambda \bar \theta}\; (\partial_{\theta}\; \partial_\lambda
\bar \eta + \varepsilon_{\lambda \mu}\; \partial_{\theta}\; 
\partial_{\bar\theta} \Phi^{\mu})\nonumber\\
J_{\bar \theta} &=& s^{\theta\theta}\; \partial_{\bar\theta} \;
\partial_{\theta} \; \eta + \partial_{\bar \theta}\; (\partial \cdot \Phi)
- \varepsilon^{\lambda \theta} \;(\partial_{\bar\theta}\; \partial_\lambda
\eta + \varepsilon_{\lambda \mu}\; \partial_{\bar\theta}\; \partial_{\theta}
\Phi^{\mu})
\end{array} \eqno(6.4)
$$
and that of $H's$ are:
$$
\begin{array}{lcl}
H_{\rho} &=& \varepsilon_{\rho\sigma}
\partial^{\sigma} \;(\varepsilon^{\mu\nu} \partial_{\mu}  \Phi_{\nu}) 
+ \varepsilon^{\lambda \bar \theta}\; \varepsilon_{\rho\sigma}\;
\partial^\sigma \bigl (\;\partial_\lambda\;
\eta -  \partial_{\bar\theta}\; \Phi_{\lambda} \bigr )
- s^{\theta\theta} \;\varepsilon_{\rho\sigma}\; \partial^\sigma
\;\partial_{\theta}\; \bar \eta \nonumber\\
&-& s^{\bar\theta\bar\theta}\; \varepsilon_{\rho\sigma}
\partial^{\sigma}\; \partial_{\bar\theta}\; \eta
+ \varepsilon^{\lambda \theta}\; \varepsilon_{\rho\sigma}
\partial^{\sigma}\; \bigl (\;\partial_\lambda
\bar \eta -  \partial_{\theta}\; \Phi_{\lambda} \bigr )
- s^{\theta\bar\theta}\;\varepsilon_{\rho\sigma} \partial^\sigma
\;\bigl (\;\partial_{\bar\theta} \bar \eta 
+ \partial_{\theta}\; \eta \bigr )\nonumber\\
H_{\theta} &=& s^{\theta\theta}\; \partial_{\bar\theta} \;
\partial_{\theta} \;\bar \eta - \partial_{\bar\theta}\; 
(\varepsilon^{\mu\nu} \partial_{\mu} \Phi_{\nu})
- \varepsilon^{\lambda \theta}\; (\partial_{\bar\theta}\; \partial_\lambda
\bar \eta -  \partial_{\bar\theta}\; \partial_{\theta}
\Phi_{\lambda})\nonumber\\
H_{\bar \theta} &=& s^{\bar\theta\bar \theta}\; \partial_{\theta} \;
\partial_{\bar\theta} \; \eta - \partial_{\theta}\; 
(\varepsilon^{\mu\nu} \partial_{\mu}  \Phi_{\nu})
- \varepsilon^{\lambda \bar \theta} \;\bigl (\partial_{\theta}\; 
\partial_\lambda \eta -  \partial_{\theta}\; 
\partial_{\bar\theta} \Phi_{\lambda} \bigr ).
\end{array} \eqno(6.5)
$$
In the computations of (6.4) and (6.5), we have exploited the anti-symmetry-
and symmetry properties of $\varepsilon^{\mu\theta(\bar\theta)}$ and
$s^{\theta\bar\theta}$ respectively.  We have also used: 
$(\partial_{\theta})^2 = 0, (\partial_{\bar\theta})^2  = 0$, etc.

Now we impose the analogue of the horizontality condition. In fact,
the condition $ d \theta\; (J_{\theta} + H_{\theta}) = 0$ leads to setting
the coefficients of $s^{\theta\theta}, s^{\bar \theta \bar \theta},
\varepsilon^{\mu\theta}, \varepsilon^{\mu \bar \theta}$ equal to zero. These
are 
$$
\begin{array}{lcl}
\bar s = 0 \;\;\; \partial_{\theta} (\partial \cdot \Phi)
= \partial_{\bar \theta} (\varepsilon^{\mu\nu}\partial_{\mu} \Phi_\nu)\;\;\;
\partial_{\theta} (\partial_{\mu} \bar \eta + \varepsilon_{\mu\nu}
\partial_{\bar \theta} \Phi^\nu) = 0 \;\;\;
 \partial_{\bar \theta} (\partial_{\mu} \bar \eta -
\partial_{ \theta} \Phi_\mu) = 0. 
\end{array}\eqno(6.6) 
$$
In the language of the component fields of the expansion (3.2), these
conditions lead to
$$
\begin{array}{lcl}
\partial \cdot S = 0 \quad \varepsilon^{\mu\nu} \partial_{\mu} S_{\nu} = 0
\quad \partial \cdot \bar R = \varepsilon^{\mu\nu} \partial_\mu R_\nu \quad
S_\mu = \partial_\mu B \quad S_\mu = \varepsilon_{\mu\nu} \partial^\nu \bar
{\cal B} \end{array} \eqno(6.7)
$$
which imply the results that are listed below
$$
\begin{array}{lcl}
\bar R_\mu = \varepsilon_{\mu\nu} R^\nu \quad \Box S_\mu = 0 \quad
S_\mu = \frac{1}{2} [\; \partial_\mu B + \varepsilon_{\mu\nu} \partial^\nu
\bar {\cal B}\; ] \quad \partial_\mu B - \varepsilon_{\mu\nu} \partial^\nu
\bar {\cal B} = 0.  \end{array} \eqno(6.8)
$$
It will be noticed that conditions $\partial \cdot S = 0$ and
$ \varepsilon^{\mu\nu}
\partial_{\mu} S_\nu = 0$ imply $\Box S_\mu = 0$. We also obtain  $\Box B = 0, 
\Box \bar {\cal B} = 0$.  Similarly, the imposition of the constraint
$ d \bar \theta \;(J_{\bar \theta} + H_{\bar \theta}) = 0$, leads to 
$$
\begin{array}{lcl}
s = 0\; \;\; \partial_{\bar \theta} (\partial \cdot \Phi)
= \partial_{\theta} (\varepsilon^{\mu\nu}\partial_{\mu} \Phi_\nu)\;\;\;
\partial_{\bar \theta} (\partial_{\mu}  \eta + \varepsilon_{\mu\nu}
\partial_{ \theta} \Phi^\nu) = 0 \;\;\;
 \partial_{\theta} (\partial_{\mu} \eta -
\partial_{ \bar \theta} \Phi_\mu) = 0. 
\end{array}\eqno(6.9) 
$$
The above equations, vis-a-vis expansions in (3.2), lead to
$$
\begin{array}{lcl}
\partial \cdot S = 0 \quad \varepsilon^{\mu\nu} \partial_{\mu} S_{\nu} = 0
\quad \partial \cdot  R = \varepsilon^{\mu\nu} \partial_\mu \bar R_\nu \quad
S_\mu = - \partial_\mu \bar B \quad S_\mu = - \varepsilon_{\mu\nu} 
\partial^\nu {\cal B}. \end{array} \eqno(6.10)
$$
These equations imply the following results
$$
\begin{array}{lcl}
 R_\mu = \varepsilon_{\mu\nu} \bar R^\nu \quad \Box S_\mu = 0 \quad
S_\mu = - \frac{1}{2} [\; \partial_\mu \bar B 
+ \varepsilon_{\mu\nu} \partial^\nu
{\cal B}\; ] \quad \partial_\mu \bar B - \varepsilon_{\mu\nu} \partial^\nu
{\cal B} = 0  \end{array} \eqno(6.11)
$$
which, in turn, lead to $ \Box \bar B = 0, \Box {\cal B} = 0$. {\it It will be
interesting to point out that the conditions $R_\mu = \varepsilon_{\mu\nu}
\bar R^\nu, \bar R_\mu = \varepsilon_{\mu\nu}  R^\nu$
are satisfied for both the decompositions (5.2) as well as (5.5)}.
The final requirement due to the analogue of
horizontality condition is
$$
\begin{array}{lcl}
d x^\rho\; (J_{\rho} + H_{\rho}) = d x^\rho\; \Box A_{\rho}. 
\end{array} \eqno(6.12)
$$
A close look at the expressions for $J_\rho$, $H_{\rho}$ and (6.12)
shows that
$$
\begin{array}{lcl}
 d x^\rho \bigl [ \partial_{\rho} (\partial \cdot \Phi) 
+ \varepsilon_{\rho\sigma}
\partial^\sigma (\varepsilon^{\mu\nu} \partial_{\mu} \Phi_{\nu}) \bigr ]
=\; d x^\rho \Box A_{\rho} 
\end{array} \eqno(6.13)
$$
where the l.h.s. reduces to $ (d x^\rho \Box \Phi_\rho) $. 
This leads to the following restrictions on  
$R_\mu, \bar R_{\mu}, S_\mu$:
$$
\begin{array}{lcl}
\Box R_{\mu} = 0\;\; \;\qquad\;\; \;\;\Box \bar R_{\mu} = 0\; \qquad\;\; 
\;\;\;\;\Box S_{\mu} = 0.
\end{array}\eqno(6.14)
$$
Hereafter, in the rest of computations, we set $ s(x) = 0, \bar s(x) = 0$ 
in the expansion (3.2) (cf.(6.6),(6.9)). Now setting the coefficients of 
$(dx^\rho s^{\theta\theta})$ and $(dx^\rho s^{\bar \theta \bar \theta})$
equal to zero leads to:
$$
\begin{array}{lcl}
 \partial_{\theta} (\partial_{\rho} \eta - 
\varepsilon_{\rho\sigma} \partial^\sigma \bar \eta ) = 0 \quad
 \partial_{\bar \theta} (\partial_{\rho} \bar \eta - 
\varepsilon_{\rho\sigma} \partial^\sigma  \eta ) = 0.  
\end{array}\eqno(6.15)
$$
The above equations produce the following two restrictions: 
$$
\begin{array}{lcl}
\partial_\rho B - \varepsilon_{\rho\sigma} \partial^\sigma {\cal B} = 0
\qquad
\partial_\rho \bar B - \varepsilon_{\rho\sigma} \partial^\sigma 
\bar {\cal B} = 0.
\end{array}\eqno(6.16)
$$
It will be noted that these restrictions automatically satisfy the 
conditions that emerge due to setting the coefficient of 
$(d x^\rho s^{\theta\bar \theta})$ equal
to zero. A comparison between (6.8) and (6.11) on the one hand and (6.16) 
on the other hand, produces the conditions: $ B = \bar B, {\cal B} 
= \bar {\cal B}, \partial_\rho B - \varepsilon_{\rho\sigma} 
\partial^\sigma {\cal B} = 0$.  However,
with these equalities, it is clear that equations (6.8) and (6.11) lead to:
$S_\mu = 0, \partial_\rho B + \varepsilon_{\rho\sigma} 
\partial^\sigma {\cal B} = 0$.
Thus, we have got the orthogonal combinations of derivatives 
on $B$ and ${\cal B}$
that are equal to zero. Ultimately, these imply that we can choose: 
$ B = \bar B ={\cal B} = \bar {\cal B} = 0$. At this stage, 
it is worthwhile to
emphasize that, as the operator equations, these conditions are {\it not}
unexpected. In fact, it has been shown that the choice of the physical state
as the harmonic state in the Hodge decomposition allows one to require
that: $ Q_{a(b)} |phys> = 0, Q_{a(d)} |phys> = 0$ which imply that
$ B |phys> = 0, {\cal B} |phys> = 0$ (cf.(7.1)) [1,2,6]. There is one more 
subtle point that has to be stressed here. Even though, $S_\mu = 0$ in terms
of $B$ and ${\cal B}$, there is another choice of $S_\mu$ that remains
invariant under the discrete symmetry transformations (2.9). For instance,
$S_\mu = - \varepsilon_{\mu\nu} \Box A^\nu$ remains
invariant under the duality transformations (2.9) because $A_\mu
\rightarrow A_\mu$ and $ \Box \rightarrow \Box$ under
$\partial_\mu \rightarrow \pm i \varepsilon_{\mu\nu} \partial^\nu$.
Furthermore, the conditions $\partial \cdot S = 0$ and
$\varepsilon^{\mu\nu} \partial_\mu S_\nu = 0$ are automatically satisfied
because $\Box (\partial \cdot A) = 0$ and $\Box E = 0$ due to the
equations of motion (with $ B = - (\partial \cdot A), {\cal B} = E$).
Thus, we shall persist with $S_\mu$ in our further discussions.

For the rest of our computations, we shall set $ s(x) = \bar s(x) = 0$
and $ B(x) = \bar B(x) = {\cal B}(x) = \bar {\cal B}(x) = 0$ in our
superfield expansion (3.2). With these insertions in (3.2), it is 
obvious now that setting of the coefficients of 
$(dx^\rho \varepsilon^{\lambda\theta})$ and 
$(dx^\rho \varepsilon^{\lambda\bar\theta})$ equal to zero, leads to
$$
\begin{array}{lcl}
\varepsilon_{\rho\sigma} \partial^\sigma \bigl (\partial_\lambda 
\bar \eta - \partial_{\theta} \Phi_\lambda \bigr ) &-& \partial_\rho
\bigl (\partial_{\lambda} \eta + \varepsilon_{\lambda\mu}
\partial_\theta \Phi^\mu) = 0 \nonumber\\
\varepsilon_{\rho\sigma} \partial^\sigma \bigl (\partial_\lambda 
\eta - \partial_{\bar \theta} \Phi_\lambda \bigr ) &-& \partial_\rho
\bigl (\partial_{\lambda} \bar \eta + \varepsilon_{\lambda\mu}
\partial_{\bar \theta} \Phi^\mu) = 0.
\end{array}\eqno(6.17)
$$
In terms of the component fields of (3.2), the above equations produce
the following conditions:
$$
\begin{array}{lcl}
\partial_\lambda \bigl ( \varepsilon_{\rho\sigma} \partial^\sigma \bar C
- \partial_\rho C \bigr ) &=& \partial_\rho  R_\lambda + 
\varepsilon_{\rho\sigma} \partial^\sigma \bar R_\lambda \nonumber\\
\partial_\lambda \bigl ( \varepsilon_{\rho\sigma} \partial^\sigma  C
- \partial_\rho \bar C \bigr ) &=& \partial_\rho  \bar R_\lambda + 
\varepsilon_{\rho\sigma} \partial^\sigma  R_\lambda \nonumber\\
\varepsilon_{\rho\sigma} \partial^\sigma S_\lambda + \partial_\rho
(\varepsilon_{\lambda\mu} S^\mu) &=& 0
\end{array}\eqno(6.18)
$$
where $R_\mu = \varepsilon_{\mu\nu} \bar R^\nu, \bar R_\mu =
\varepsilon_{\mu\nu} R^\nu$. The last equation, contracted with 
$\partial_\lambda$, is automatically satisfied due to 
$(\partial\cdot S) = 0$ and $\varepsilon^{\mu\nu} \partial_\mu S_\nu = 0$.
Moreover, the top equation can be re-expressed, in terms of the
component fields, as
$$
\begin{array}{lcl}
- \partial_\lambda \bigl ( \varepsilon_{\rho\sigma} \partial^\sigma  C
- \partial_\rho \bar C \bigr ) = \partial_\rho  \bar R_\lambda + 
\varepsilon_{\rho\sigma} \partial^\sigma  R_\lambda.
\end{array}\eqno(6.19)
$$
Comparison of the central equation of (6.18) with the above equation, yields
$$
\begin{array}{lcl}
 \partial_\lambda \bigl ( \varepsilon_{\rho\sigma} \partial^\sigma  C
- \partial_\rho \bar C \bigr ) = 0 \qquad \partial_\rho  \bar R_\lambda + 
\varepsilon_{\rho\sigma} \partial^\sigma  R_\lambda = 0.
\end{array}\eqno(6.20)
$$
One of the very interesting solutions for the above conditions is
$\bar R_\rho = \partial_\rho \bar C - \varepsilon_{\rho\sigma} 
\partial^\sigma C = 0$ which is equivalent to setting the Hodge
decomposed version of equation (5.2) equal to zero. This, in turn,
implies that the other 2D fermionic vector $R_\rho = \partial_\rho C
- \varepsilon_{\rho\sigma} \partial^\sigma \bar C$ is also equal to
zero. Thus, ultimately, we have obtained the following solution
from the analogue of the horizontality condition w.r.t. the super
Laplacian operator $\tilde \Delta$
$$
\begin{array}{lcl}
R_\mu (x) &=& \bar R_\mu (x) = 0 
\qquad B(x) = \bar B(x) = {\cal B}(x) = \bar {\cal B}(x) = 0
\nonumber\\
S_\mu (x) &=& - \varepsilon_{\mu\nu}\; \Box \; A^\nu (x) \equiv
\bigl ( \tilde s_{w} A_\mu (x) \bigr ) \qquad s(x) = \bar s(x) = 0.
\end{array}\eqno(6.21)
$$
The superfield expansion (3.2), with the above inputs, bears an appearance
 as
$$
\begin{array}{lcl}
\Phi_\mu (x, \theta, \bar \theta) = A_\mu (x) + i\; \theta \bar\theta\;
(\tilde s_{w} A_{\mu}(x)) \qquad \eta (x, \theta, \bar \theta) = C(x) \qquad
\bar \eta (x, \theta, \bar \theta) = \bar C(x).
\end{array}\eqno(6.22)
$$
This shows that the ghost fields $C(x)$ and $\bar C(x)$ do not transform
at all under $\tilde s_{w}$ and gauge field $A_\mu(x)$ transforms to its own
equation of motion along the bosonic ($\theta\bar\theta$) direction
of the expansion. 
This establishes the fact that symmetry transformations 
$\tilde s_{w} C = 0, \tilde s_{w} \bar C = 0, \tilde s_{w} A_{\mu} 
= - \varepsilon_{\mu\nu} \Box A^\nu$, which are the analogues of (2.7) 
(with $ {\cal B} = E, B = 
- (\partial \cdot A))$, are generated by
a conserved bosonic charge $Q_{w}$ (cf.(7.1)) which owes its origin
to the super Laplacian operator $\tilde \Delta$. Geometrically, the
operator $Q_{w}$ turns out to be the translation generator along a bosonic
direction that is equivalent to
a couple of successive translations along $\theta$ and $\bar \theta$
directions (i.e. $\mbox{Lim}_{\theta,\bar\theta\; \rightarrow 0}\;
\frac{\partial}{\partial \theta}\; \frac{\partial}{\partial \bar \theta}\;
\Phi_\mu (x,\theta,\bar\theta) \;=\; [A_\mu, Q_{w}]$).\\

\noindent
{\bf 7 Conclusions}\\

\noindent
In the present investigation, we have exploited the 
mathematical power of super de Rham cohomology
operators ($ \tilde d, \tilde \delta, \tilde \Delta $) of differential geometry
and a generalized version of the so-called horizontality condition 
[13-18] for the geometrical interpretation of 
the local Noether conserved charges $(Q_{(a)b}, Q_{(a)d}, Q_{w})$
as the generators of translations along the Grassmannian (odd)- and 
bosonic (even)
directions of a compact $(2 + 2)$-dimensional supermanifold.
The explicit {\it local} expressions for all these conserved 
Noether charges, together with the ghost charge $Q_{g}$, are [1,2,6]
$$
\begin{array}{lcl}
Q_{b} &=& {\displaystyle \int d x}\; \bigl [\; B \dot C 
- \dot B C\; \bigr] \;\qquad
Q_{ab} = {\displaystyle \int d x}\; \bigl [\; B \dot {\bar C} 
- \dot B \bar C \;\bigr] \nonumber\\
Q_{d} &=& {\displaystyle \int d x}\; \bigl [\; {\cal B} \dot {\bar C} 
- \dot {\cal B} \bar C \; \bigr] \;\qquad
Q_{ad} = {\displaystyle \int d x}\; \bigl [\; {\cal B} \dot  C 
- \dot {\cal B}  C \;\bigr] \nonumber\\
Q_{w} &=& {\displaystyle \int d x}\; \bigl [ \;B \dot {\cal B}  
- \dot B {\cal B}  \;\bigr] \;\qquad
Q_{g} = - i {\displaystyle \int d x }\;
\bigl [\; C\; \dot {\bar C} + \bar C\; \dot C \;\bigr ]
\end{array}\eqno(7.1) 
$$
which obey the following algebra [1,2,6]
$$
\begin{array}{lcl}
&& Q_{b}^2 = Q_{d}^2 = Q_{ab}^2 = Q_{ad}^2 = 0 \;\;\qquad
 \{ Q_{b},   Q_{ab} \} = \{ Q_{d}\; Q_{ad} \} = 0 
\nonumber\\
&& i\; [\;  Q_{g}, Q_{b(ad)}\; ] = +\; Q_{b(ad)} \;\;\qquad \;\;
 i \;[\;  Q_{g}, Q_{d(ab)}\; ] = - \;Q_{d(ab)} \nonumber\\
&&Q_{w}  = \{ Q_{b}, Q_{d} \} = \{ Q_{ad}, Q_{ab} \} \;\qquad
 [Q_{w}, Q_{k}] = 0\; \quad k = a, ab, d, ad, g
\end{array}\eqno(7.2)
$$
if we exploit the canonical (anti-)commutation relations for the Lagrangian
density (2.2). The above algebra is reminiscent of the algebra obeyed by the
de Rham cohomology operators of differential geometry (see,e.g.,Sec.1) and
it remains form-invariant under the discrete duality transformations (2.9).
The degree of a given differential form is equivalent to the ghost number 
of a quantum state in the BRST formalism as (1.1) and (7.2) imply 
$$
\begin{array}{lcl}
i Q_{g} Q_{b(ad)}\; |\Psi>_n &=& (n + 1)\; Q_{b(ad)}\; |\Psi>_n \nonumber\\
i Q_{g} Q_{d(ab)}\; |\Psi>_n &=& (n - 1)\; Q_{d(ab)} \;|\Psi>_n \nonumber\\
i Q_{g} Q_{w} \;|\Psi>_n &=& n \; Q_{w}\; |\Psi>_n.
\end{array}\eqno(7.3) 
$$
The above equations re-establish the identification of conserved local 
Noether charges 
with the de Rham cohomology operators of differential geometry
because as operators $d(\delta)$ increase(reduce) the degree of a form by one
so do the local charges $Q_{b(ad)}(Q_{d(ab)})$ to a state with a certain ghost
number. Thus, one has the following mappings:
$ d \Leftrightarrow (Q_{b}, Q_{ad}), \delta \Leftrightarrow
(Q_{d}, Q_{ab}), \Delta \Leftrightarrow Q_{w} = \{Q_{(a)b}, Q_{(a)d} \}$.
With these identifications, the HDT and the BRST cohomology can be defined 
comprehensively in the quantum Hilbert space of states [1,2,6,11].
The existence of some new discrete symmetries in the superfield formulation 
leads to some key insights which enable
us to obtain the Hodge decomposition for the 2D odd vectors of 
the superfield expansion (3.2). The 
interplay between discrete and continuous symmetries of the theory (together
with the idea of the HDT) provides the origin for the existence
of (anti-)BRST- and (anti-) co-BRST symmetries.
The requirement of the self-duality invariance of the theory under (2.9)
provides a logical reason for the choice of the Hodge decomposed version
(5.2) over its counterpart (5.5) in the solution (6.20). In fact, this
solution finally leads to the derivation of the bosonic symmetry where
$R_\mu = \bar R_\mu = 0$. The self-duality invariance also provides
a very convincing guess for the correct Hodge decomposed version of
2D bosonic vectors: $S_\mu = \partial_\mu B + \varepsilon_{\mu\nu} \partial^\nu
{\cal B}$ and $ A_\mu = \partial_\mu \lambda + \varepsilon_{\mu\nu}
\partial^\nu \kappa $ where $\lambda$ and $\kappa$ are some scalar fields
(see, e.g., Ref. [2] for details).

In our earlier works [6,1,2], it has been demonstrated that 2D free Abelian
$U(1)$ gauge theory in the flat Minkowski spacetime is a new type of 
topological field theory (TFT) [23] as the BRST- and dual BRST symmetries gauge
out the propagating degrees of freedom of the 2D photon. This new
theory captures together some of the key features of Witten- as well as 
Schwarz type TFTs [24,25]. In fact, the form of the quantum action for this
theory bears the appearance of a Witten type theory but the symmetries are
local gauge type which resemble very much like symmetries of the 
Schwarz type TFT. The
topological nature of this theory has been shown to be encoded in the
vanishing of the Laplacian operator $Q_{w} \rightarrow 0$ when the on-shell
condition for photon (i.e. $ \partial_{\mu} B + \varepsilon_{\mu\nu}
\partial^\nu {\cal B} = 0$) is utilized [1,2,6]. In fact, under this condition,
all the transformations in (2.7) become trivially zero. In the framework of
superfield formulation, this fact is encoded in the vanishing of the 2D
fermionic vectors $R_\mu$ and $\bar R_\mu$ and bosonic 2D vector $S_\mu$
turns out to be equal to $-\varepsilon_{\mu\nu} \Box A^\nu$ {\it only}.
Thus, on the on-shell (i.e. $\Box A_\mu = 0$), even $S_\mu$ turns out to 
be zero.  The topological nature of the 2D free $U(1)$ gauge theory is
captured in the horizontality condition of the superfield formulation. In
fact, it can be seen that the restriction
$\tilde \Delta \tilde A = \Delta A (\equiv dx^\rho \Box A_\rho)$ vanishes on 
the on-shell ($\Box A_\mu = 0$)
of the 2D photon which, ultimately, implies $Q_{w} \rightarrow 0$. In a recent 
paper [26], a very direct, simple and straightforward method has been adopted 
to capture some of the topological features of our present theory where
an explicit presence of the superfields $\Phi_\mu (x,\theta,\bar\theta),
 \eta (x,\theta,\bar\theta)$ and
$\bar \eta (x,\theta,\bar\theta)$ has been invoked.

It is important to point out that, in the superfield formulation on the
($2+2$)-dimensional supermanifold, one can define forms that have greater
degree than the spacetime dimension of the supermanifold. However, the
generalized versions of the horizontality condition w.r.t. all the super
cohomological operators  (which we have exploited 
in our present discussion) turn out to be quite handy to resolve this
anomaly. It would be nice to take an explicit example of a supersymmetric
theory and work out all the details of the HDT for this theory in the quantum
Hilbert space of states. At the moment, we are looking into this problem
and our results will be reported in our future publications.

An interesting generalization of our present superfield formulation for the 
free 2D Abelian gauge theory to self-interacting 2D non-Abelian gauge theory
has already been studied in Ref. [27]. It would be very interesting
endeavour to look for the field theoretical models for the Hodge theory
where there is an interaction between matter- and gauge fields. Some steps
in this direction have already been taken [3,4] for the interacting 2D
Abelian gauge theory. It has been shown in these works that, as (anti-)BRST
symmetries are connected with the local gauge symmetries on gauge-
and Dirac fields,
similarly the (anti-)co-BRST symmetries on the gauge fields are connected
with the analogue of chiral symmetries on the Dirac fields. The local
Noether charges, corresponding to these symmetries, turn out to be the
analogues of cohomolgical operators $d$ and $\delta$. We firmly believe
that a thorough study of this interacting theory with the Wess-Zumino
term would shed light on the consistency and unitarity of the 2D anomalous
gauge theory (see, e.g., Refs. [28,29] and references therein) in the 
framework of BRST cohomology and HDT. It would be also interesting to
generalize our results to the free 4D two-form Abelian gauge theory where
the existence of  analogues of the local conserved charges 
$Q_{(a)b}, Q_{(a)d}, Q_{w}$ has been established. These are some of the 
issues that are under investigation [30].\\

\noindent
{\bf Acknowledgements}\\

\noindent
The enlightening and thought-provoking comments by the adjudicator on some 
of the key issues of this paper are gratefully acknowledged. Thanks are also 
due to A. Lahiri for his interest in this work.

\end{document}